\documentclass[a4]{article}
\usepackage[english]{babel}
\usepackage[utf8]{inputenc}
\usepackage[mathletters]{ucs}
\usepackage{amsmath}
\usepackage{graphicx}
\usepackage{subcaption}
\usepackage{epsfig}
\usepackage[colorinlistoftodos]{todonotes}
\usepackage[hidelinks]{hyperref}
\usepackage[margin=1in]{geometry}
\usepackage{multicol}
\usepackage{caption}[=v1]
\usepackage{lmodern}
\usepackage{siunitx}
\usepackage[super,comma,sort&compress]{natbib}

\usepackage{multirow}
\usepackage{booktabs}
\usepackage{authblk}
\usepackage{xr}
\usepackage{setspace}

\title{Chemical state detection and charge transfer in complex oxide heterostructures via \textit{in situ} Auger Electron Spectroscopy}

\date{}
\author[1]{Harish Kumarasubramanian}
\affil[1]{Mork Family Department of Chemical Engineering and Materials Science, University of Southern California, Los Angeles, CA 90089, USA }

\author[1,2,3,*]{Jayakanth Ravichandran}

\affil[2]{Ming Hsieh Department of Electrical Engineering, University of Southern California, Los Angeles, California 90089, USA}
\affil[3]{Core Center of Excellence in Nano Imaging, University of Southern California, Los Angeles, California 90089, USA}
\affil[*]{Email: j.ravichandran@usc.edu}
\graphicspath{Images_new }

\begin{document}
\maketitle

\begin{abstract}

Understanding and controlling the chemical states both in the bulk and at the interfaces of complex oxide thin films is essential for engineering a wide range of electronic, optical, and magnetic functionalities, which arise through emergent phenomena such as two-dimensional electron gases, interfacial magnetism, and associated phase transitions. Here, we demonstrate the use of \textit{in situ} Auger Electron Spectroscopy (AES) as a powerful tool for probing oxidation states and dynamic chemical processes during the growth of complex oxide heterostructures. By leveraging the chemical sensitivity of AES to subtle changes in valence electron populations, we show that this technique can distinguish distinct oxidation states in multivalent perovskite manganate and vanadate systems with high fidelity during deposition. Furthermore, we show evidence for dynamic chemical phenomena, specifically charge transfer processes at the polar–nonpolar LaMnO$_3$/SrTiO$_3$ interface. Our results establish \textit{in situ} AES as a powerful diagnostic tool for monitoring and controlling interfacial chemistry during thin film growth, offering a pathway toward the atomic-scale engineering of chemical states in functional oxide heterostructures.

\end{abstract}

\normalsize 
\section*{Introduction}

Complex oxide heterostructures exhibit extraordinary electrical, magnetic\cite{izyumskaya2009oxides}, optical\cite{sando2018epitaxial}, and multiferroic\cite{vaz2010magnetoelectric} properties, offering immense potential for a diverse array of emerging technologies\cite{lorenz20162016}. Their precise chemical composition and carefully controlled oxidation states are critical for applications in advanced energy-efficient electronics\cite{ramesh2024roadmap}, optical communication systems\cite{karvounis2020barium}, multifunctional sensors, and a wide range of other technologies that exploit novel interfacial phenomena\cite{mannhart2010oxide}. Although many of these oxide-based applications are still in research or prototype stages, their successful transition to practical technologies is strongly dependent on achieving unprecedented compositional accuracy and interfacial control of composition and chemistry at the atomic scale.

Achieving atomic-scale precision in thin film deposition demands powerful \textit{in situ} characterization tools capable of monitoring and controlling growth dynamics as they occur. In recent decades, structural monitoring techniques integrated directly into deposition chambers have significantly improved thin-film growth quality. In particular, Reflection High-Energy Electron Diffraction (RHEED) has emerged as a critical \textit{in situ} structural characterization method, widely adopted for the growth of complex oxide thin films even at high pressures\cite{rijnders1997situ}. RHEED has enabled continuous, sub-monolayer-sensitive monitoring of surface crystallinity, growth modes, surface reconstructions, and thickness evolution, facilitating breakthroughs in achieving atomically sharp interfaces and epitaxial control necessary for high-performance oxide-based devices.

\begin{figure}
 \centering\includegraphics[width=0.8\linewidth]{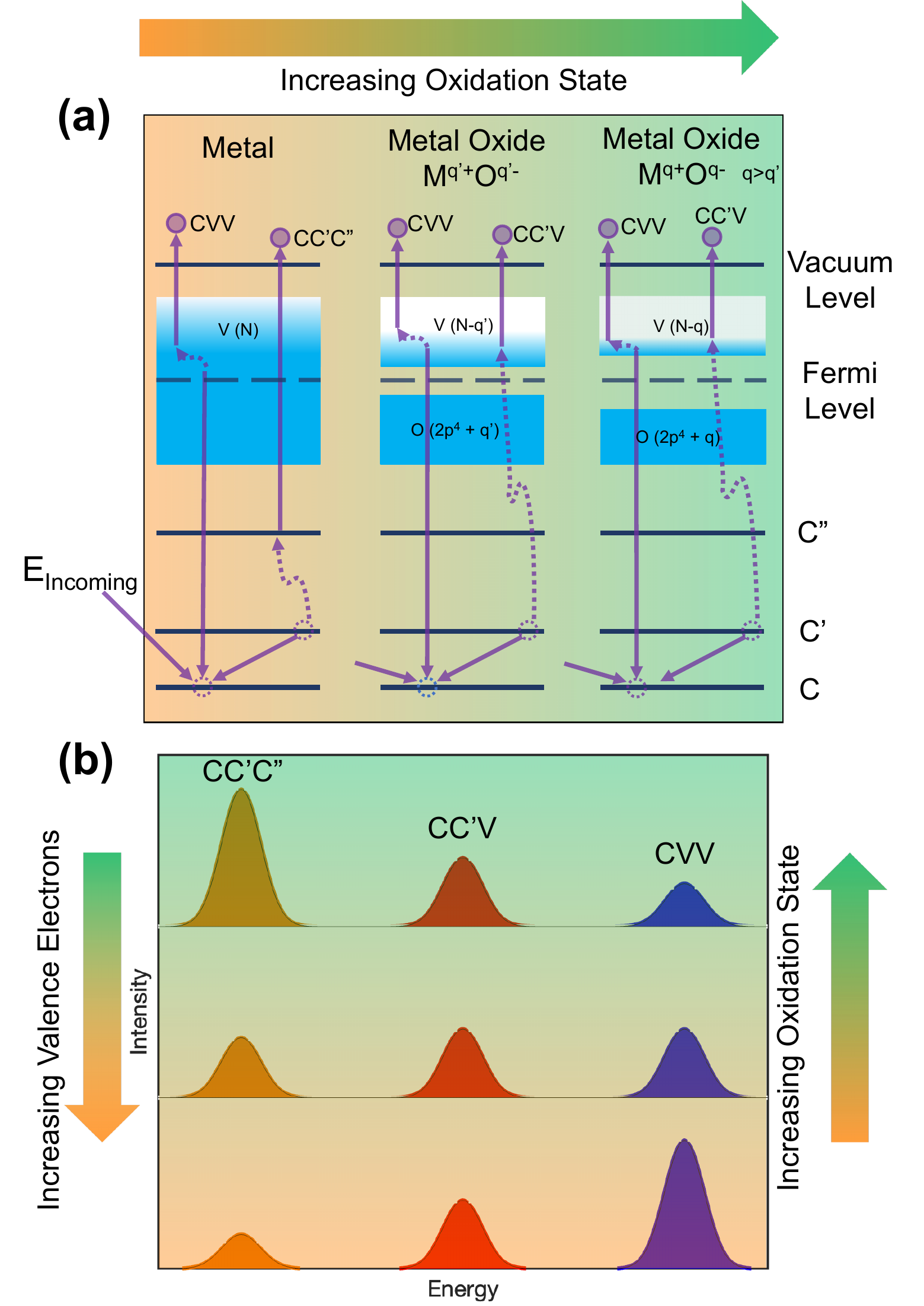}
 \captionof{figure}{ (a) Electron occupation in the valence band of metal and metal oxides as a function of oxidation states. Core shells of increasing binding energy are marked as C'', C' and C. Auger transitions involving these core levels and the valence band are shown. For the metal oxides, the electronic charge lost to oxygen is shown as q' and q, wherein $q'<q$. N is the number of valence electrons or corresponds to the electron population in the valence band of a nascent metal. The valence electron population in the metal oxide after oxidation is (N-q') and (N-q) respectively. The difference in the shades of the valence bands at different oxidation states indicates this. Valence electron population decreases progressively as the metal is increasingly oxidized. (b) Qualitative variation of the corresponding Auger fine spectra of type CXX' for the metal and metal oxide scenarios shown in (a). C is a Core level and X and X' are either core or valence levels. The intensities of Auger transitions that involve valence electrons (CVV and CC'V) decreases as the metal is increasingly oxidized.}
  \label{fig:LMM spectra}
\end{figure}

Despite these successes in structural characterization, \textit{in situ} chemical and compositional analysis remains comparatively underdeveloped, often limited to \textit{ex situ} or ``\textit{pseudo in situ} methods" involving sample transfer and exposure to altered environmental conditions. Despite multiple reports of chemical state measurement and charge transfer in various oxide heterostructures\cite{comes2016interface,comes2017probing,thapa2021probing,kleibeuker2014electronic,mahatara2022high} through \textit{in situ} methods, those that probe these chemical features directly during growth at the same temperature, pressure, and growth environment have been lacking. Ultimately, \textit{in situ} approaches that involve an in-vacuo sample transfer might not capture the complete picture of transient surface phenomena, sensitive compositional fluctuations, and critical oxidation states for heterostructures of all chemistries, particularly those sensitive to the growth environment. Consequently, subtle but decisive factors that influence growth dynamics shall remain hidden during growth, hampering the precise control of functional properties of these heterostructures.

Addressing these limitations requires integrating truly \textit{in situ}, chemically sensitive analytical techniques directly into thin-film deposition environments, with a potential for real time operation. Recent advances in Auger Electron Spectroscopy (AES)\cite{staib2011situ}, which is inherently surface-sensitive and capable of probing compositional changes at the atomic scale, have enabled its successful integration into elevated oxygen pressures that are typical in oxide thin film deposition methods such as Pulsed Laser Deposition (PLD)\cite{orvis2019situ}. It has recently been demonstrated that the use of such an \textit{in situ} AES probe allows real-time monitoring of the film composition, particularly precise identification and quantification of surface termination\cite{orvis2021situ} and, crucially, also leads to active control and manipulation of termination during film deposition\cite{orvis2021direct}.

The development of a real-time, \textit{in situ} chemical state characterization using Auger Electron Spectroscopy (AES) represents a significant advancement in monitoring thin-film growth. This advance leverages the intrinsic sensitivity of AES to surface chemical composition and electronic states. Historically, the capability of AES to detect chemical states emerged prominently in the late 1970s and early 1980s, driven by improved spectral resolution for higher-energy Auger transitions in inorganic systems. Early foundational studies demonstrated that the fine structure and relative intensity ratios of Auger peaks, particularly transitions involving outer-shell electrons, are highly responsive to changes in valence electron configurations, and thus, oxidation states. Specifically, Allen \textit{et al.}\cite{allen1977high} established a direct correlation between the intensity ratios of ``LMM" Auger transitions in the first-row transition metals and their respective number of \textit{d}-electrons. Extending this insight, Rao \textit{et al.}\cite{rao1980novel} further demonstrated similar correlations in oxide systems, revealing how variations in the oxidation state markedly influence Auger spectra. The relations developed and verified by them are shown in equation \ref{eq:AugerIntensities}. Figure \ref{fig:LMM spectra} (a) highlights how the valence electron population evolves with the oxidation state: as oxidation state increases, the valence band is progressively depleted, reducing the likelihood that valence electrons participate in the Auger process. Consequently, the  relative intensities of the transitions vary qualitatively as illustrated in figure \ref{fig:LMM spectra}(b), wherein Auger transitions in the first and second group transition metals are of the type CXX'. Here, C represents a core shell and X and X' denote either core or valence shells, with the intensity changes directly reflecting the valence electron density.

\begin{equation}
       \frac{I_{CVV}}{I_{CC'C'}} \propto (N-q)(N-q-1)\,, \quad\quad  \frac{I_{CC'V}}{I_{CC'C'}} \propto (N-q)\,
       \label{eq:AugerIntensities}
\end{equation} 

Here, C and C' are the core levels, and V is the valence level. $I_{xyz}$ represents the intensity for Auger electrons that transition from $x\rightarrow y\rightarrow z$ energy levels. N and q are the number of valence electrons in the neutral atom and the nominal oxidation state, respectively. Subsequently, N-q would represent the number of electrons remaining in the valence level. Thus, the relative intensities of auger transitions with different valence electron contributions would vary in a monotonic fashion with the oxidation state.

In this article, we employ \textit{in situ} AES to monitor the oxidation states of epitaxial complex oxide heterostructures of two different elemental systems deposited using PLD. In addition to distinguishing the chemical states in the bulk, we also use the depth-sensitive nature of AES to show the evolution of chemical states during growth, demonstrating the acquisition of nearly real-time chemical signatures of interfacial phenomena such as charge transfer in these systems. By combining the escape depth of Auger electrons and the sensitivities of their relative intensities to the oxidation states, a parameter-free model is built and is used to quantitatively compare the experimental results. Leveraging this chemical sensitivity \textit{in situ} and in real-time provides researchers the unique ability to monitor and precisely control oxidation states and chemical composition during thin-film deposition processes, thus enabling unprecedented atomic-scale chemical control in complex oxide thin-film growth.

\section*{Results and Discussion}

To demonstrate the chemical sensitivity of the Auger probe for distinguishing oxidation states, perovskite complex oxides of manganese and vanadium were selected as model systems due to their well-known ability to exhibit multiple oxidation states. LaMnO$_3$ (LMO) and CaMnO$_3$ (CMO) were chosen to represent Mn in nominal +3 and +4 oxidation states, respectively. Similarly, LaVO$_3$ (LVO) and SrVO$_3$ (SVO) were selected to represent the +3 and +4 oxidation states of vanadium.

\subsection*{Chemical Sensitivity in the bulk: Bilayers}

For 1st row transition metals, C, C' and V in equation \ref{eq:AugerIntensities} are the L$_{23}$,  M$_{23}$ and M$_{45}$ energy levels, respectively. These are referred to as the L, M and V energy levels respectively (for brevity) in figure \ref{fig:nomspectra}(a) and (b) and the spectra together is called the ``LMM'' spectra in the rest of the article. As a first experiment to assess the method’s sensitivity to bulk chemical state variations independent of depth resolution, bilayers consisting of 100 unit cells (UCs) each of CMO and LMO, and 100 unit cells each of SVO and LVO were grown independently on (001)-oriented SrTiO$_3$ (STO) substrates. During the deposition of each bilayer, Mn and V LMM Auger spectra were acquired \textit{in situ} at intervals of every 20 UCs. The representative Mn and V LMM spectra, along with the oxygen KLL peaks for each of the manganate and vanadate films, are shown in figure \ref{fig:nomspectra}(a) and (b) respectively. Given that the mean free paths of Mn and V LMM Auger electrons are estimated to be in the range of 1–1.3 nm\cite{seah1979quantitative}, each 20 UC increment (approximately 8 nm) is sufficient to effectively attenuate any signal from the underlying layers. This ensures that the recorded spectra are representative of the chemical state evolution of the films with $\sim$20 UC intervals, with negligible contributions from the previously deposited (and monitored) layer. Concurrently, the film growth was monitored using Reflection High-Energy Electron Diffraction (RHEED). The growth mode and rate were tracked by observing the oscillations and intensity of the specular spot in the RHEED pattern. 

\begin{figure}
 \centering\includegraphics[width=0.85\linewidth]{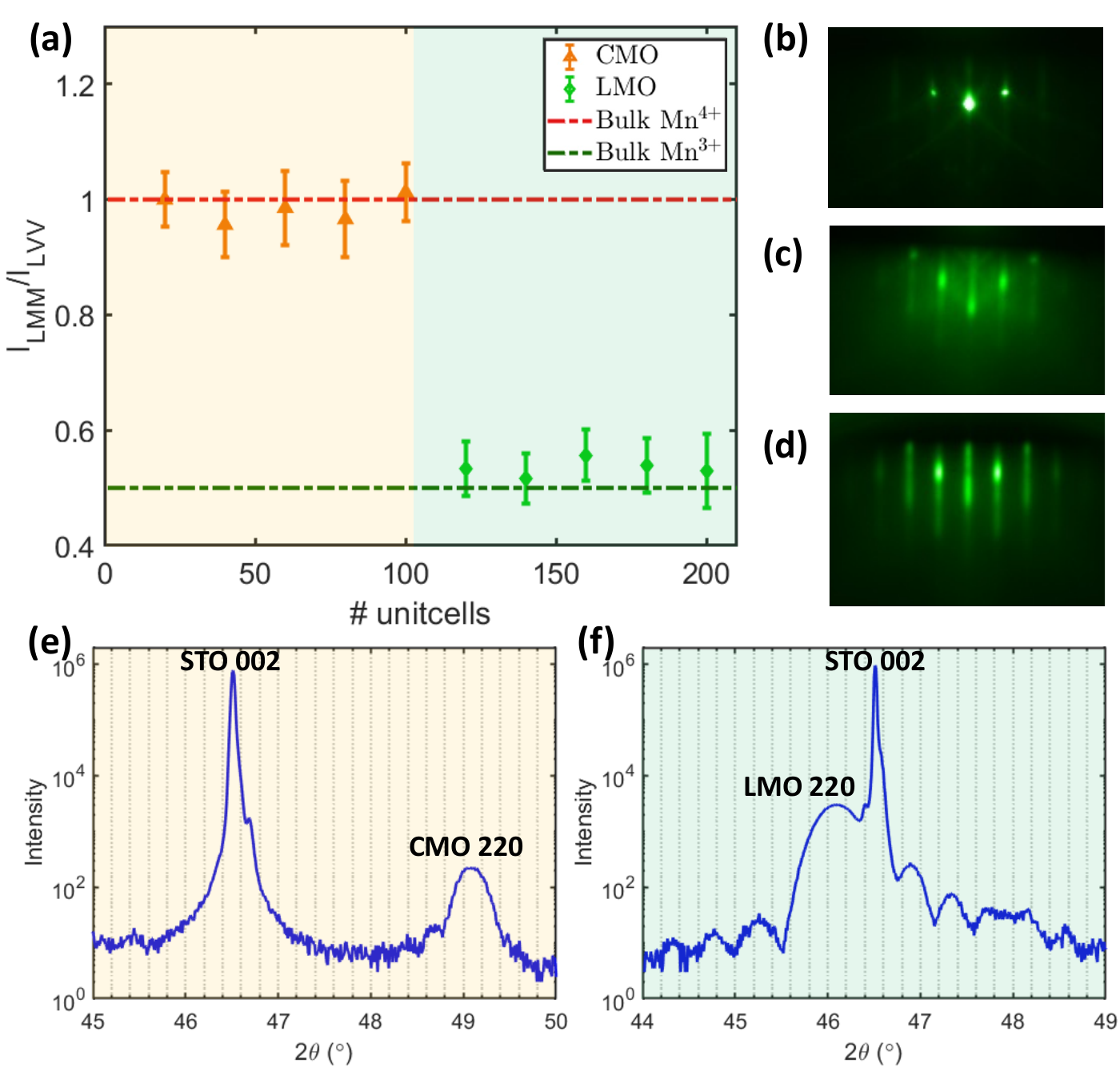}
   \captionof{figure}{(a) The ratio of intensities of the LMM and LVV peaks of Manganese in 100 UC LMO/ 100 UC CMO bilayers. The orange and green shaded regions represent CMO and LMO respectively. RHEED pattern of: (b) the SrTiO$_{3}$ (001) substrate before growth, (c) the film surface after the deposition of 100 UC of CMO, and (d) the film surface after subsequent deposition of 100 UC of LMO. High resolution out-of-plane thin film X-ray diffraction patterns of $\sim$ 20 nm thick films of (e) CMO and (f) LMO grown on STO (001) substrates. The CMO in (e) is completely relaxed while the LMO shown in (f) is strained.}
 \label{fig:Bulk Auger_Mn}
\end{figure}
 
\begin{figure}
 \centering\includegraphics[width=0.85\linewidth]{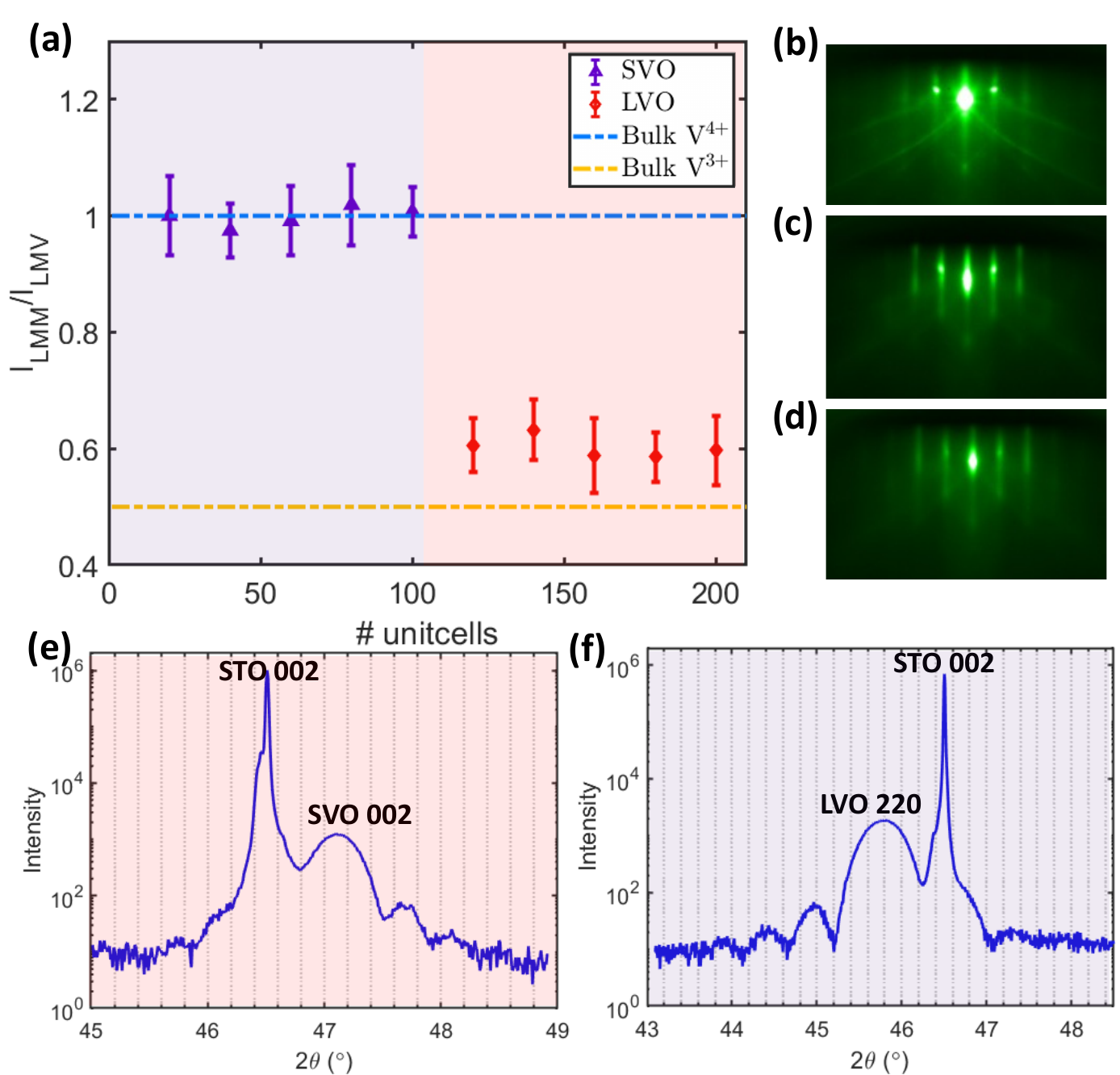}
 \captionof{figure}{(a) The ratio of intensities of the LMM and LMV peaks of Vanadium in 100 UC SVO/ 100 UC LVO bilayers. The orange and green shaded regions represent SVO and LVO respectively. RHEED pattern of: (b) the SrTiO$_{3}$ (001) substrate before growth, (c) the film surface after the deposition of 100 UC of SVO, and (d) the film surface after subsequent deposition of 100 UC of LVO. (e) High resolution out-of-plane thin film X-ray diffraction patterns of $\sim$ 20 nm thick films of SVO and LVO grown on STO (001) substrates.}
  \label{fig:Bulk Auger_V}
\end{figure}

Figures \ref{fig:Bulk Auger_Mn} (a) and \ref{fig:Bulk Auger_V} (a) show the relative Auger intensities of the ``LMM'' peaks for both Mn and V respectively in their respective bilayer configurations. Equation \ref{eq:AugerIntensities} gives the relative intensity ratios for these peaks at their respective oxidation states. Theoretical intensity ratios, calculated using equation \ref{eq:AugerIntensities}, are overlaid, and the peak intensity ratios at each 20 UC interval were normalized relative to the first data point. For manganese, the LMM and LVV transitions were analyzed. The Mn LVV peak and the La MNN peaks are separated by 6 eV and usually appear as a flat double peak. 2 gaussian curves were fit for this double peak as a whole to extract the Mn LVV and La MNN intensities separately.  In contrast, for vanadium, LVV transitions are largely suppressed in compounds wherein the vanadium oxidation states are close to +4 or higher. This is evident from both the spectra shown in figure \ref{fig:nomspectra}(b),wherein the LVV peak is nearly absent in the SVO film and from the theoretical relationship in equation \ref{eq:AugerIntensities}, where the ratio of the LVV to LMM peak intensities scales with (N–q–1). With N = 5 (number of valence electrons for nascent V) and q = 4 (oxidation state), the numerator approaches zero, rendering the intensity for the LVV transition negligible. Consequently, LMV transitions were used instead for vanadium in this analysis.

The Auger intensity ratios for the LMO/CMO bilayer shown in figure \ref{fig:Bulk Auger_Mn} (a) clearly reflect Mn in the +3 and +4 oxidation states, consistent with the expected valences of the two compounds. However, in the case of vanadium, figure \ref{fig:Bulk Auger_V} (a) shows the measured intensity ratios deviating from those expected for ideal +3 and +4 oxidation states in LVO and SVO. This discrepancy may arise due to several factors. Strain-induced valence shifts are known to occur in thin-film oxide systems where transition metals can accommodate different oxidation states \textit{via} octahedral distortions\cite{chandrasena2017strain} and oxygen vacancy formation. Since CMO has a pseudocubic lattice parameter of 3.72 $\si{\angstrom}$\cite{imbrenda2016surface,flint2014magnetism}, the lattice mismatch on STO(3.905 $\si{\angstrom}$) is large enough to cause a relaxation within a few UCs. The X-Ray Diffraction (XRD) pattern of 20 nm thick CMO on STO in figure \ref{fig:Bulk Auger_Mn}(e) reflects this, where the CMO 220 reflection has a $2\theta$ of $49^{\circ}$ which corresponds to the bulk lattice parameter of 3.72 $\si{\angstrom}$. Hence, the LMO/CMO bilayers are fully relaxed. Both LVO and SVO have pseudocubic lattice parameters of 3.925 $\si{\angstrom}$ and 3.84 $\si{\angstrom}$, and so the vanadate bilayers could remain strained over a thickness of 100 UCs or so.\cite{fouchet2016study,moyer2013highly} Hence, the deviation in the oxidation state from the bulk could arise presumably from the strain effect. XRD patterns of 20 nm thick films of SVO and LVO are shown in figure \ref{fig:Bulk Auger_V} (e) and (f) respectively. The out-of-plane lattice parameters extracted from these scans, 3.85 $\si{\angstrom}$ and 3.96 $\si{\angstrom}$ for SVO and LVO, deviate slightly from the expected strained out-of-plane lattice constants of 3.83 $\si{\angstrom}$\cite{moyer2013highly} and 3.94 $\si{\angstrom}$\cite{cheikh2024investigation} respectively, and agree with this line of reasoning. Past reports have shown that highly reducing atmospheres employed during the growth of the SVO\cite{fouchet2016study} and LVO shall lead to non-stoichiometry, especially oxygen vacancies, thereby both reducing the vanadium oxidation state and expanding the lattice. 
Given that the intensity ratios from the ``LMM" Auger transitions provide relative, rather than absolute, measures of the oxidation state, it is challenging to pinpoint the exact valence state of vanadium in each film at this stage. Nevertheless, the clear difference in relative peak intensities between LVO and SVO layers confirms that AES can reliably distinguish between different oxidation states \textit{in situ}, and \textit{quantitatively} in the bulk.

\subsection*{Depth sensitive oxidation state evolution}

We next explore our ability to monitor the depth-sensitive evolution of chemical states \textit{in situ} for both manganate and vanadate systems. One key advantage of Auger electrons is their short inelastic mean free paths (IMFP) (approximately 1-1.5 nm for the Mn and V LMM peaks involved\cite{seah1979quantitative}), making AES extremely sensitive to surface compositional and chemical variations. This intrinsic sensitivity is particularly beneficial for tracking chemical state changes across thin layers or interfaces as they form during growth. To validate this capability, heterostructures of 20 UC CMO/ 20 UC LMO and 20 UC SVO/ 20 UC LVO were independently deposited on TiO$_2$-terminated STO (001) substrates with Mn and V LMM spectra measured \textit{in situ} every few unit cells. Charge transfer processes have been extensively documented at the interfaces of polar/non-polar perovskite oxide heterostructures\cite{hwang2012emergent,xu2016quasi,moetakef2011electrostatic}, where polar discontinuities are commonly compensated by charge transfer and changes in valence state of transition metal ions near the interface. The LaAlO$_3$/SrTiO$_3$ (LAO/STO) system is a prototypical example of an oxide heterostructure demonstrating charge transfer at the interface arising from the polar discontinuity between the polar LAO film and the nonpolar STO substrate\cite{singh2011built}. Similarly, charge transfer phenomena have been extensively reported at other interfaces, such as LaMnO$_3$/SrTiO$_3$\cite{chen2017electron,mundy2014visualizing,wang2015imaging,li2019controlling} and LaVO$_3$/SrTiO$_3$\cite{hotta2007polar,balal2017electrical,halder2022unconventional,tomar2020conducting}, due to analogous polar discontinuities ((Mn/V)O$_2^{-1}$/LaO$^{+1}$/TiO$_2$/SrO). 

\subsubsection*{Manganates on STO}

\begin{figure}
 \centering\includegraphics[width=0.78\linewidth]{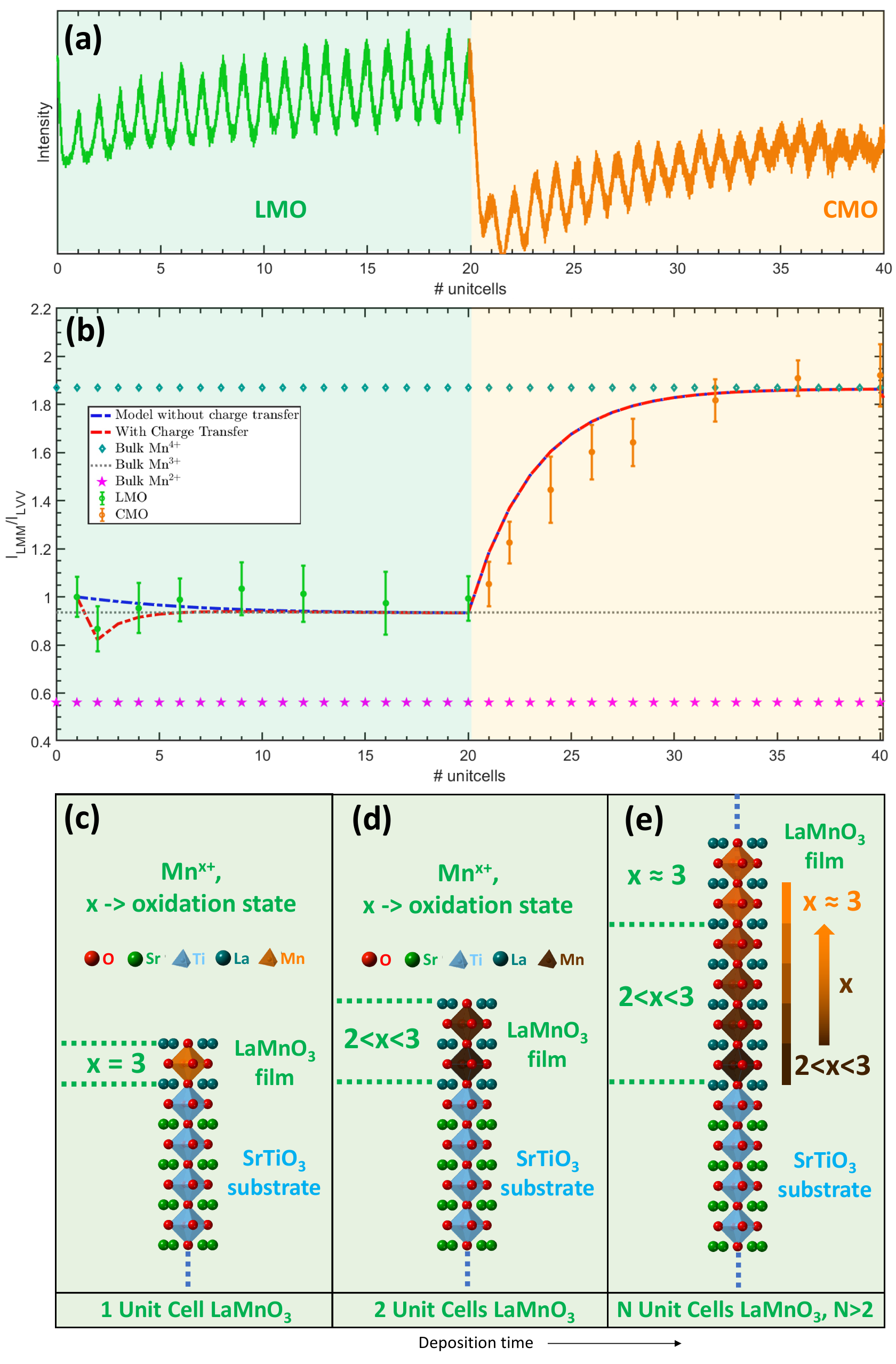}
   \captionof{figure}{(a) Time dependent evolution of the intensity of the specular spot during the growth of LMO and CMO layers indicating layer-by-layer growth. (b) The ratio of LMM and LVV intensity values for 20 UC CMO/ 20 UC LMO heterostructure grown on STO substrate. The Bulk +4,+3 and +2 oxidation state levels of Mn are marked for reference. Parameter free escape depth models with and without charge transfer in LMO/STO interface are plotted. The green and orange shaded regions represent LMO and CMO respectively. Oxidation states of Mn after the deposition of (c) 1 unit cell, (d) 2 unit cells - the advent of charge transfer and (e) $>$ 2 unit cells. The shades of the MnO$_6$ octahedra qualitatively represent the oxidation state in Mn, with it gradually increasing to +3 as the LMO layer gets thicker.  }
 \label{fig:Chargetransfer_Mn}
\end{figure}

For the LMO/STO interface, previous studies have shown that the electrostatic potential arising from the polar discontinuity triggers a charge transfer after the deposition of approximately two unit cells of LMO\cite{chen2017electron,kaspar2019electronic,wang2015imaging}, with Mn close to the interface adopting a lower oxidation state. However, nearly all such spectroscopic evidence to date has been obtained via \textit{ex situ} techniques: either by analyzing multiple samples grown separately with varying thicknesses, or through post-growth cross-sectional Electron Energy Loss Spectroscopy (EELS) imaging performed in Scanning Transmission Electron Microscopy (STEM)\cite{mundy2014visualizing,oberaigner2023visualising,kaspar2019electronic} or via `pseudo \textit{in situ}' techniques mentioned earlier.

Here, we employ our chemical analysis capability of \textit{in situ} AES to track the evolution of the Mn oxidation state directly during the growth of 20 UC CMO / 20 UC LMO layers. Figure \ref{fig:Chargetransfer_Mn}(b) shows the intensity ratios of the Mn LMM to LVV Auger transitions during the overall 40-unit-cell growth. Spectra were collected after deposition of 1, 2, 4, 6, 8, 12, 16, and 20 unit cells within each of the LMO and CMO layers, with intensity ratios normalized to the first unit cell of LMO. The nominal Mn LMM/LVV intensity ratios corresponding to bulk Mn valence states of +4, +3, and +2 are marked for reference.

We first focus on LMO growth. A pronounced dip in the LMM/LVV intensity ratio is observed after deposition of two unit cells of LMO, indicating that interfacial charge transfer lowers the Mn valence below the bulk-like +3 state. With further deposition, the measured ratio gradually approaches the bulk-like +3 value. To analyze the thickness dependence quantitatively, we employed a parameter-free escape-depth model (described in the supplementary information). Two scenarios: one assuming no interfacial charge transfer and another incorporating charge transfer are compared in Figure \ref{fig:Chargetransfer_Mn}(b). A simplified charge-transfer model that assumes a reduced Mn valence of +2 at the interface (first unit cell), followed by +3 from the second unit cell onward, captures the dip observed at 2 UC reasonably well.

Importantly, the reduced ratio at 2 UC should be interpreted as the signature of a depth-dependent charge distribution rather than a single, uniquely defined interfacial valence. After deposition of 2 UC, the Mn valence is expected to be lowest closest to the interface and to increase monotonically toward the surface, and the measured ratio represents an escape-depth-weighted average over this near-surface region. As deposition proceeds, the surface-sensitive measurement progressively samples Mn farther from the interface, and the data indicate that by approximately after the 4th unit cell of LMO, the surface region has nominally returned to the bulk-like +3 value. This evolution is schematically illustrated in Figure \ref{fig:Chargetransfer_Mn}(c)–(e).

It is also important to note that the drop in the LMM/LVV ratio can only confirm the presence of \textit{some} Mn$^{2+}$ character near the interface; multiple charge distributions can fit the observed trend. Figure \ref{fig:LMOchargedist} presents plausible charge distributions across the first two unit cells, including gradual valence changes (e.g., Mn valence 2.7 in the first UC increasing to 2.8 in the second UC) that match the data more closely. A continuous charge distribution in which the Mn valence gradually approaches +3 as the LMO layer thickens is therefore more likely than a strictly discrete (+2 then +3) scenario. However, because the number of possible distributions increases rapidly with the number of layers over which charge is spread, we restrict the modeling to charge distributions confined to the first two unit cells for modeling feasibility.

Finally, the depth resolved charge distribution inferred at a given thickness is not necessarily identical to that inferred at earlier thicknesses, as the chemical potential does evolve as the layers are being deposited. This implies that the depth-resolved oxidation state inferred after 2 UC growth need not be the same as that after 4 UC growth, and that once 20 UC of LMO is deposited, the charge distribution near the buried interface is likely different from that measured at these early stages.

Next, we consider the chemical evolution during the deposition of CMO, where the change in oxidation state from +3 to +4 is much more prominent. Initially, the Auger signal reflects contributions from both oxidation states: +3 from the underlying LMO layer and +4 from the newly deposited CMO layer. As the CMO thickness increases, the contribution from LMO is progressively attenuated, and beyond approximately 10 UC the ratio saturates, predominantly reflecting the bulk-like +4 oxidation state of CMO with minimal contributions from LMO. The RHEED oscillations shown in Figure \ref{fig:Chargetransfer_Mn}(a) were used as a guide to acquire the Auger spectra precisely at the completion of a layer.

Several additional points emerge from these measurements and modeling efforts. First, the LMM/LVV ratio is below unity in the no-charge-transfer model as the LVV electrons have a larger mean free path than LMM electrons (13.6~\si{\angstrom} for LVV vs. 12.6~\si{\angstrom} for LMM) and therefore saturate over a longer distance. Second, minor deviations from the bulk +3 oxidation state at larger thicknesses ($>$5 UCs) may indicate subtle strain effects arising from lattice mismatch with the STO substrate. Third, the slower-than-expected increase in Mn valence during the initial CMO layers may reflect strain-relaxation before 10 UCs \cite{pedroso2020strain}, which could lower the effective oxidation state and/or alter the effective IMFP values from the bulk values used here. However, given the limits on the precision of the AES measurements, attributing this deviation to a single mechanism is difficult.

The O and Mn spectra for the LMO and CMO layers are shown in Figure \ref{fig:LMOandCMOspectra}(b) and (c), respectively. The area under the curve (AUC) of these spectra is plotted as the relative ratio in Figure \ref{fig:Chargetransfer_Mn}(b). The stoichiometry of LMO and CMO was also evaluated using the Mn LMM/La MNN peak ratio for LMO and the Mn LMM/Ca LMM peak ratio for CMO. The corresponding spectra and extracted cation stoichiometries are shown in Figures \ref{fig:LaMnStoich} and \ref{fig:CaMnStoich}. The La/Mn ratios were fitted to an escape-depth model assuming a MnO$_2^{-1}$ termination; La-excess scenarios were modeled as those leading to gradual termination changes. The data indicate that the LMO grown is not La rich. Modeling Mn-rich scenarios is more complicated because, with a starting MnO$_2^{-1}$ termination, there would be no termination change and, without an additional layer of different Mn stoichiometry, the evolution of Mn signal is not uniquely diagnostic of Mn excess. Nevertheless, XRD of LMO on STO (Figure \ref{fig:Bulk Auger_Mn}(f)) shows an out-of-plane lattice constant of 3.94~\si{\angstrom}, consistent with prior reports of strained, stoichiometric LMO on STO \cite{chen2017electron}. Moreover, Mn-rich LMO has been reported to suppress charge transfer at the STO interface \cite{chen2017electron}, supporting the conclusion that the LMO grown in this work is most likely stoichiometric.

Across all Auger data shown, while individual AUCs can vary non-monotonically with thickness (e.g., between 2 and 4 UCs in CMO in Figure \ref{fig:LMOandCMOspectra}(c)), the relative ratios between peaks robustly follow the oxidation-state trend in Figure \ref{fig:Chargetransfer_Mn}(b). Such variations in absolute intensity are expected due to small changes in accelerating voltage or emission current during deposition, which typically scale all peaks together, whereas the relative peak ratios capture the underlying chemical state and stoichiometry.

\subsubsection*{Vanadates on STO}

A similar charge-transfer phenomenon has also been observed at the LaVO$_3$/SrTiO$_3$ (LVO/STO) interface \cite{hotta2007polar,balal2017electrical,halder2022unconventional,tomar2020conducting}, with theoretical predictions and experimental evidence indicating a critical thickness of around 5 unit cells (UC). To verify this behavior, Vanadium Auger spectra were acquired at precisely matched thickness intervals for both LVO and SVO layers as previously mentioned in the manganate case. The corresponding LMM-to-LMV intensity ratios for these layers are presented in Figure \ref{fig:LVOSVO_CT}. Analogous to the manganate case, an escape-depth model was employed to fit the data. Here, the charge-transfer scenario assumes a reduced oxidation state of +2 in the first interfacial unit cell, which rapidly transitions to +3 for all the subsequent layers ($\geq $5 UC). The resulting model predicts only a marginal decrease in intensity ratios relative to a scenario without charge transfer. Specifically, at a thickness of 5 UC, the contribution of the interfacial layer is expected to be less than 10$\%$ of the total signal, with the change in the intensity-ratio due to charge transfer limited to approximately 3$\%$. These subtle changes lie below the detection threshold of the current experimental setup and thus remain unresolved. The experimentally measured intensity ratios across different thicknesses within the LVO layers show minimal variation. However, a clear and pronounced increase in the measured oxidation state is observed upon subsequent deposition of the SVO layers. This signal saturates again beyond approximately 10 UC, analogous to the behavior observed in the manganate bilayers. The deviation from the bulk oxidation states at larger thicknesses is equivalent to the scenario in the thick bilayers discussed earlier.

\section*{Summary and Outlook}

In this work, we have demonstrated the capability of \textit{in situ} Auger Electron Spectroscopy (AES) to directly measure the oxidation state and its evolution in complex oxide thin films during their deposition, providing close to real-time insight into chemical-state dynamics at the atomic level. We used the relative intensity ratios of the Auger fine spectra of transition metals to monitor distinct oxidation states in multivalent perovskite manganate and vanadate systems. Specifically, at the LaMnO$_3$/SrTiO$_3$ interface, we have presented compelling evidence of interfacial charge transfer. The ability to monitor oxidation states and subtle changes in valence electron populations in real-time, directly within the growth environment, offers unique insights over conventional \textit{ex situ} methods giving us the ability to ultimately control and engineer chemical states \textit{in situ} during the deposition of these heterostructures. 

Moreover, the chemical sensitivity and potential real-time feedback provided by this technique offers exciting prospects for extending these insights beyond complex oxide interfaces including III-V, II-VI semiconductors, TMDCs, metal-semiconductor, and metal-organic interfaces, optimizing high-performance transistors, novel optoelectronic devices, and revealing precise chemical strategies to control charge injection, barrier formation, and device stability. Additionally, such \textit{in situ} techniques allow researchers to rapidly iterate and optimize growth parameters in these systems, dramatically shortening development cycles, and significantly enhancing device reliability and reproducibility, ultimately enabling precise, atomic-level control to tailor electronic, magnetic, optical, and electrochemical properties across a wide spectrum of transformative technologies.

\section*{Methods}
\subsection*{Thin Film Growth}

 All layers were deposited on TiO$_2$ terminated-single crystal SrTiO$_3$ (001) substrates using Pulsed Laser Deposition (PLD). The LaMnO$_3$ and CaMnO$_3$ thin films were deposited at a laser fluence of 1.5 J/cm$^2$, at a repetition rate of 2 Hz and at an oxygen partial pressure of 1.65 x 10$^{-2}$ mbar at  700$^{\circ}$C from stoichiometric targets of the same materials. Subsequently, the films were cooled to room temperature at 5$^{\circ}$C/min under 100 mbar oxygen pressure. The LaVO$_3$ and SrVO$_3$ thin films were deposited at a laser fluence of 2 J/cm$^2$, at a repetition rate of 2 Hz and at an oxygen partial pressure of 8 x 10$^{-6}$ mbar at  700$^{\circ}$C from  LaVO$_4$ and Sr$_2$V$_2$O$_7$ targets respectively. Subsequently, the films were cooled to room temperature at 5$^{\circ}$C/min at the growth pressure. The target-substrate distance used for all layers was 75 mm. The films were continuously monitored \textit{situ} via Reflection High Energy Electron Diffraction.

\subsection*{Auger Electron Spectroscopy}

Auger spectra were collected \textit{in situ} using a Staib Auger Probe equipped with its dedicated software suite. Electron excitation was provided by an electron gun operating at a grazing-incidence geometry, set at an excitation voltage of 5 kV and an emission current of 5 µA. The same electron source was also employed for simultaneous acquisition of the Reflection High-Energy Electron Diffraction (RHEED) signal. To enhance data quality and reduce noise, each Auger spectrum reported represents the average of five consecutive scans, with the associated standard deviation reflecting the variability among these scans. The spectra were plotted as E·N(E) vs. E, where E denotes electron energy and N(E) is the corresponding measured intensity. For peak analysis, a linear background was fitted and subsequently subtracted from each peak, and the intensity of each peak was computed as the integrated area under the curve of width 0.5 eV.

\subsection*{X-Ray Diffraction}

The High Resolution thin film XRD scans in the out of plane and off axis geometry were carried out on a Bruker D8 Advance diffractometer using a Ge (004) two bounce monochromator at Cu K$\alpha _{1}$ ($\lambda$=1.5406 $\si{\angstrom}$) radiation. 

 \section*{Acknowledgements}
 This acquisition of the AES probe was supported by an Air Force Office of Scientific Research grant no. FA9550-16-1-0335. H.K. acknowledges USC Annenberg Fellowship. The authors gratefully acknowledge the use of facilities at the Core Center for Excellence in Nano Imaging at the University of Southern California, where X-ray diffraction (XRD) studies were conducted. H.K. and J.R. specifically thank Dr. Thomas Orvis for technical assistance and valuable input during preliminary experimental planning and discussions. The authors also gratefully acknowledge technical support from Staib Instruments, as well as the benefit of discussions with Dr. Philippe G. Staib, Dr. Eric Dombrowski, and Laws Calley.

\newpage
\setcounter{figure}{0}
\Huge 
\begin{center}
    \maketitle{Supporting Information}
\end{center}
\renewcommand\thefigure{S\arabic{figure}}

\normalsize

\onehalfspacing

\section{LMM Auger Spectra of Mn and V}

\begin{figure*}[!h]
\centering
\includegraphics[width = \textwidth]{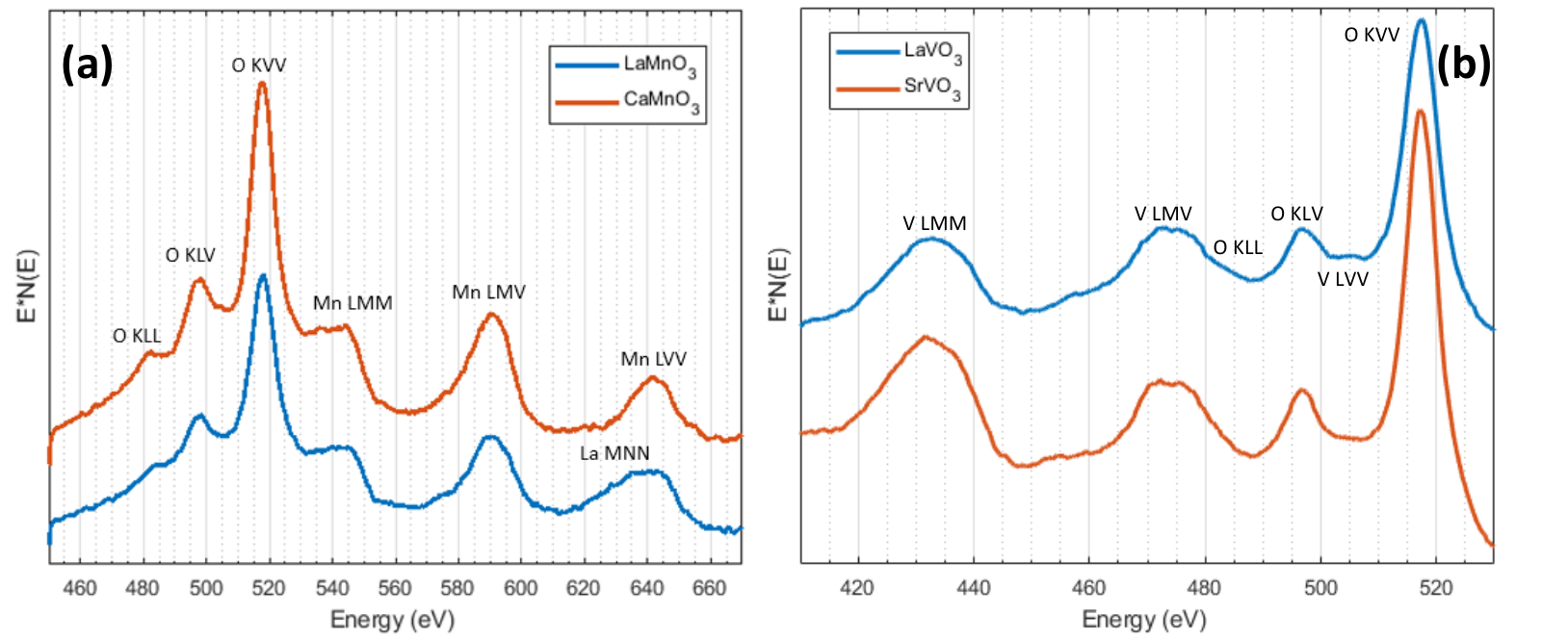}
\vspace{-0.2in}
\caption{ (a) LMM fine spectra of Mn in CMO and LMO also showing the KLL type peaks of Oxygen (b) LMM fine spectra in V in SVO and LVO also showing the KLL type peaks of Oxygen}
\label{fig:nomspectra}
\end{figure*}

\section{Parameter-free escape depth Model}

\subsection{Escape depth}

The escape depth of auger electrons for inorganic compounds was experimentally fit by Seah and Dench\cite{seah1979quantitative} as follows:

\begin{equation}
       \lambda_{MFP} =  { \frac{143}{E^2}} + 0.054\sqrt E 
       \label{eq:Auger1}
\end{equation} 

wherein E is the Energy of the Auger Electrons and $\lambda_{MFP}$ is the mean-free path of the Auger electron at Energy E. This is the model used to calculate the inelastic mean free paths of auger electrons.

\begin{figure}[h]
 \centering\includegraphics[width=0.6\linewidth]{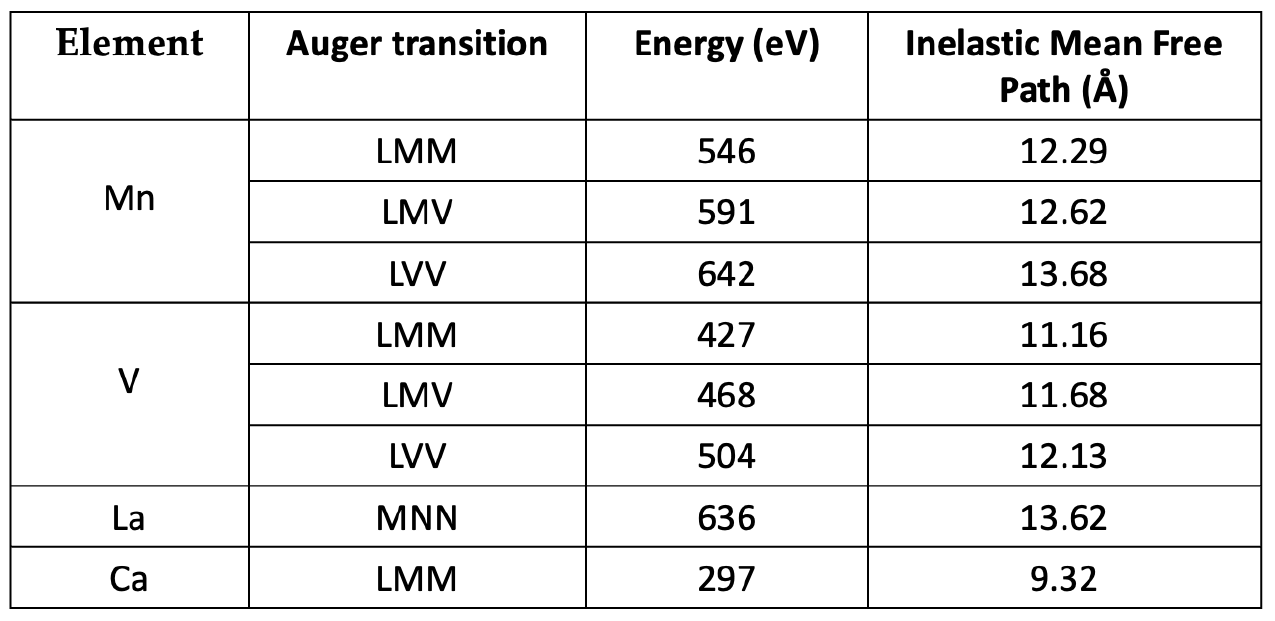}
 \captionof{figure}{Inelastic Mean Free Path values for the different auger electrons used in this study.}
 \label{fig:IMFP}
\end{figure}

\subsection{Cationic Stoichiometry}

\begin{figure}[h]
 \centering\includegraphics[width=\linewidth]{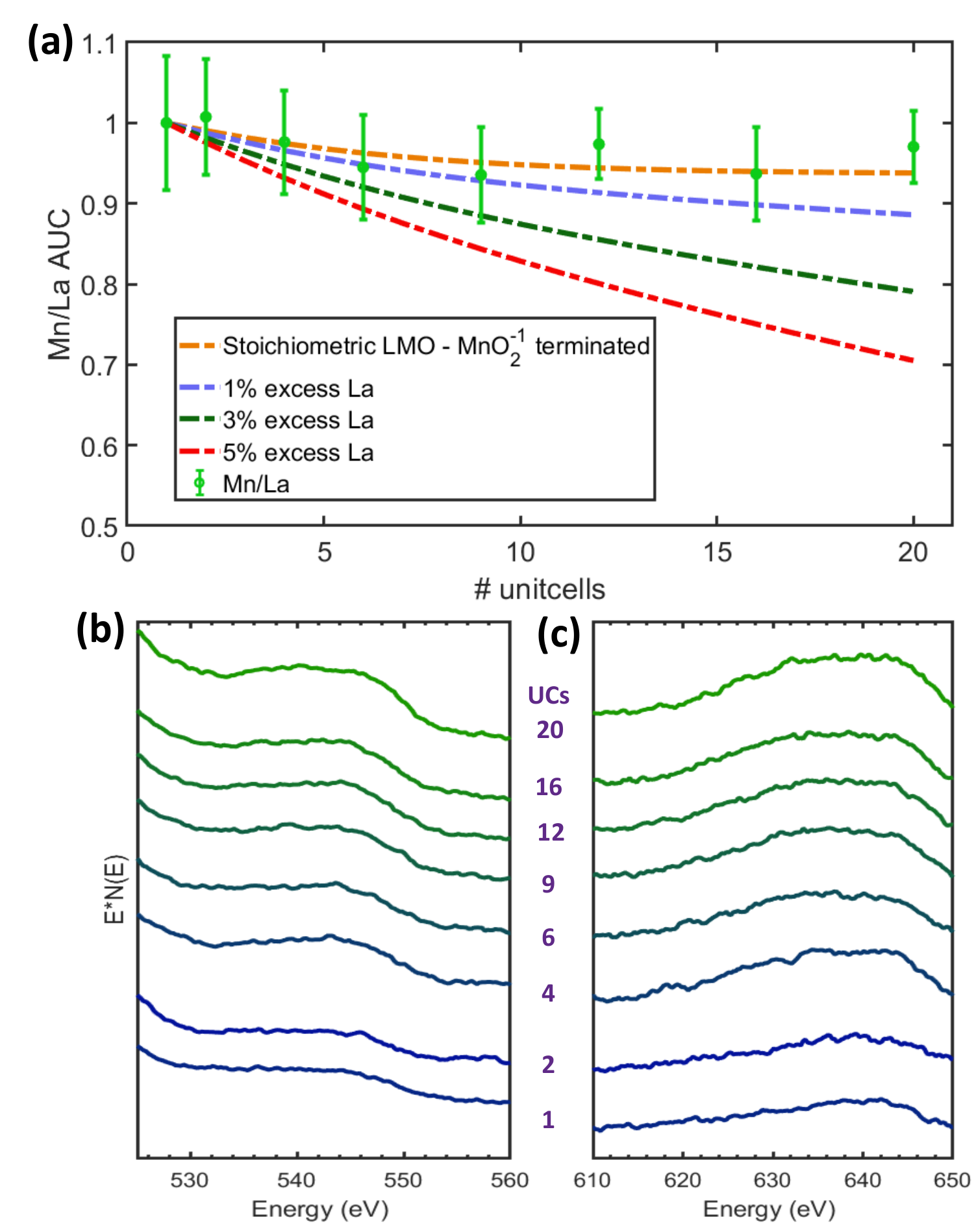}
 \captionof{figure}{ (a) Relative ratios of Mn LMM and La MNN peaks fit to a parameter free escape depth model showing a stoichiometric MnO$_2^{-1}$-terminated case along with different La-rich non-stoichiometries (b) Mn LMM and (c) La MNN peaks measured at different LMO thicknesses}
 \label{fig:LaMnStoich}
\end{figure}

\begin{figure}[h]
 \centering\includegraphics[width=\linewidth]{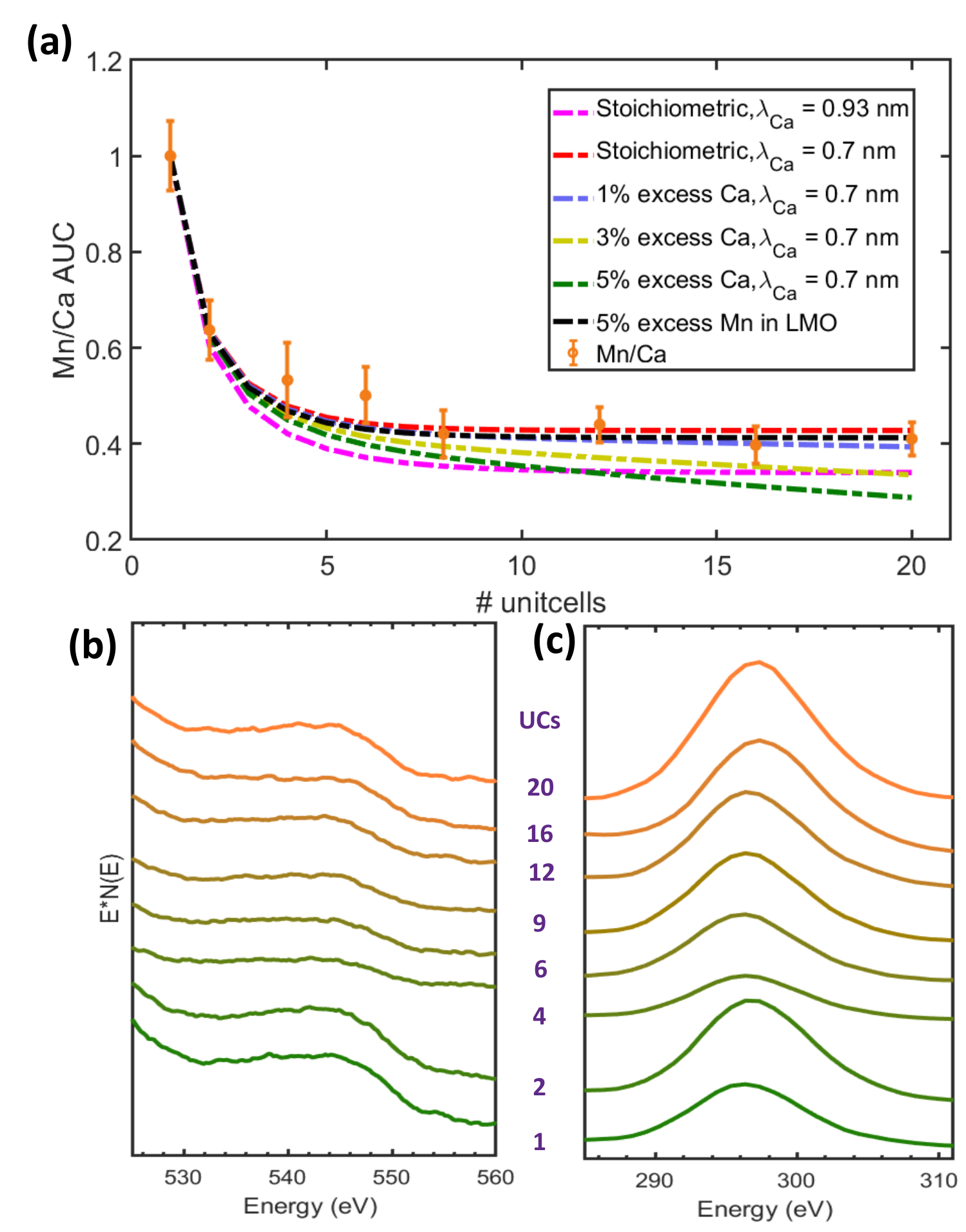}
 \captionof{figure}{(a) Relative ratios of Mn LMM and Ca LMM peaks fit to a parameter free escape depth model showing a stoichiometric MnO$_2^{-1}$-terminated case along with different Ca-rich non-stoichiometries (b) Mn LMM and (c) Ca LMM peaks measured at different CMO thicknesses}
 \label{fig:CaMnStoich}
\end{figure}

The intensity of a particular peak belonging to element A involving the transition $x\rightarrow y\rightarrow z$ , where x,y and z are energy levels is given by 

\begin{equation}
       I^A_{xyz} = \sum_{j=0}^{N-1} I_o e^{ \frac{-j*a}{\lambda_{xyz}}},
       \label{eq:Auger1}
\end{equation} 
 
where in, N is the number of repeating layers involving element A, a is the lattice parameter or the d-spacing parallel normal to the surface at which the element A repeats, I$_o$ is the intensity of the Auger electrons emitted from a single layer and $\lambda$ is the mean-free path of the associated Auger electrons.

Assuming the Element to be Mn, the transition to be LMM and the compound to be LMO, the equation becomes,

\begin{equation}
       I^{Mn}_{LMM} = \sum_{j=0}^{N-1} I_{Mn} e^{ \frac{-j*a_{LMO}}{\lambda^{Mn}_{LMM}}},
       \label{eq:2}
\end{equation} 
 where in $a_{LMO}$ is the lattice parameter of LMO. 

 LMO on a TiO$_2$-terminated STO (001) surface will give a MnO$_2^{-1}$ terminated layer with the LaO$^{+1}$ layer adjacent to the TiO$_2$ layer in STO. In that case tracking the intensity of both Mn and La from the surface a 20 UC LMO layer on a on a TiO$_2$-terminated STO (001) surface can be modelled as follows:

\begin{equation}
       I^{Mn}_{LMM} = \sum_{j=0}^{19} I_{Mn} e^{ \frac{-j*a_{LMO}}{\lambda^{Mn}_{LMM}}} \,, \quad\quad   I^{La}_{LMM} = \sum_{j=0}^{19} I_{La} e^{ \frac{-(j+1/2)*a_{LMO}}{\lambda^{La}_{LMM}}} ,
       \label{eq:3}
\end{equation} 

 The extra 1/2 term in the exponent for the La intensity is because the LaO$^{+1}$ is buried underneath the MnO$_2^{-1}$ layer.

Assuming the Mn to La ratio to be 1 to 1, the ratio of Mn to La would track as follows:

\begin{equation}
       \frac{I^{Mn}_{LMM}}{I^{La}_{LMM}} =  \frac{\sum_{j=0}^{19} I_{Mn} e^{ \frac{-j*a_{LMO}}{\lambda^{Mn}_{LMM}}}}{\sum_{j=0}^{19} I_{La} e^{ \frac{-(j+1/2)*a_{LMO}}{\lambda^{La}_{LMM}}} } \,, \quad \text{normalizing with respect to the first unit cell,} \quad \frac{I^{Mn}_{LMM}}{I^{La}_{LMM}} =  \frac{\sum_{j=0}^{19}e^{ \frac{-j*a_{LMO}}{\lambda^{Mn}_{LMM}}}}{\sum_{j=0}^{19}e^{ \frac{-(j+1/2)*a_{LMO}}{\lambda^{La}_{LMM}}} } 
       \label{eq:4}
\end{equation} 

\subsection{Oxidation State Models}

\subsubsection{Manganates}

\begin{figure}[h]
 \centering\includegraphics[width=\linewidth]{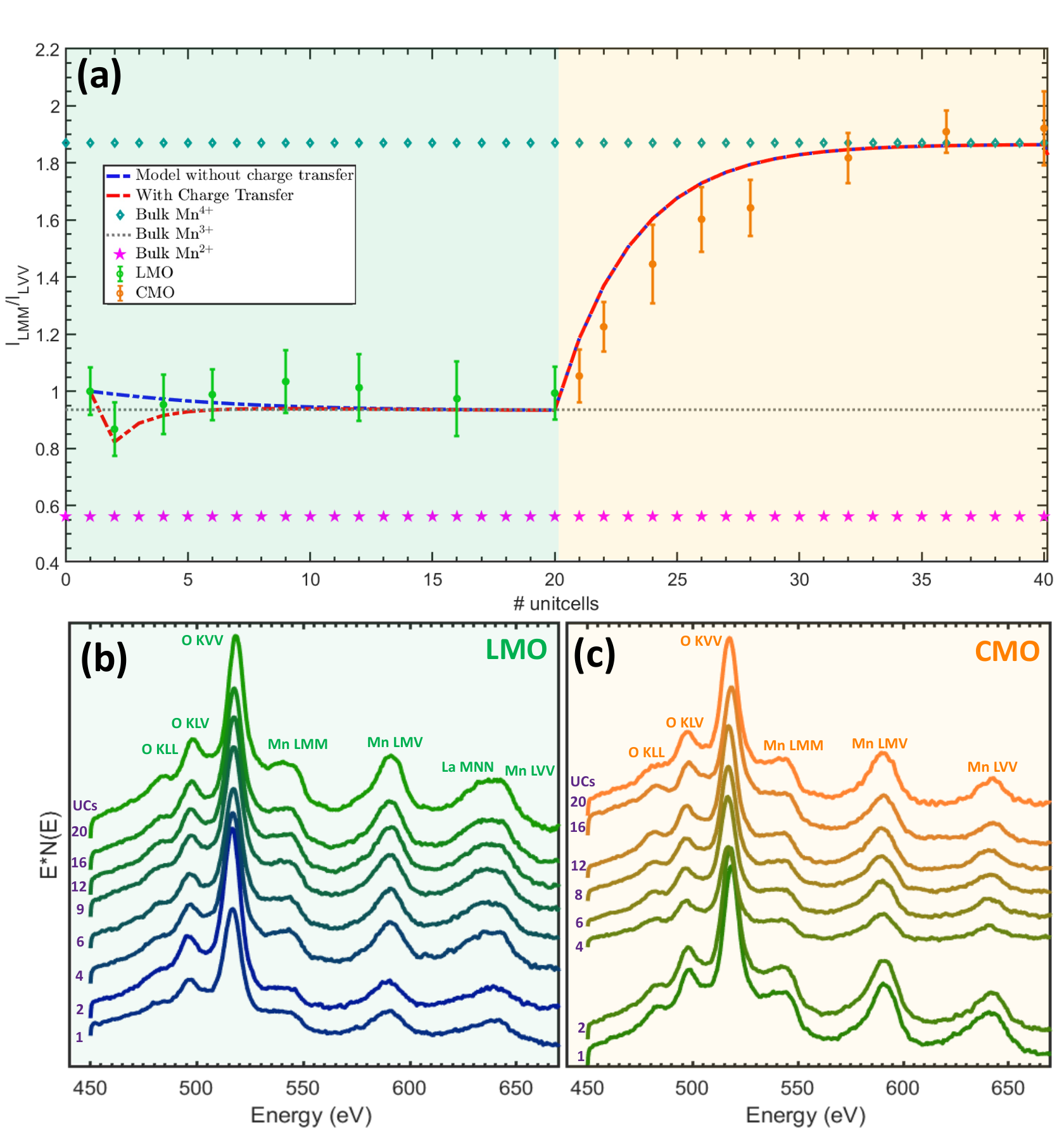}
 \captionof{figure}{(a)LMM to LVV intensity ratios in 20 UC CMO/ 20 UC LMO on STO. The Bulk +4,+3 and +2 oxidation state levels of Mn are marked for reference. Parameter free escape depth models with and without charge transfer in LMO/STO interface are plotted. The green and orange shaded regions represent LMO and CMO respectively. O and Mn spectra at different (b) LMO and (c) CMO thicknesses.}
 \label{fig:LMOandCMOspectra}
\end{figure}

For Mn, the LMM and LVV transitions have been considered to model the oxidation state evolution.

\begin{equation}
       \frac{I_{LVV}}{I_{LMM}} \propto (N-q)(N-q-1)\,, \quad \text{Wherein N is the total number of valence electrons and q is the nominal oxidation state},
       \label{eq:5}
\end{equation} 

For 20 layers of LMO with MnO$_2^{-1}$ termination and assuming all layers have Mn in the +3 oxidation state, we have

\begin{equation}
       I^{Mn}_{LVV} = \sum_{j=0}^{N} I_{Mn} e^{ \frac{-j*a_{LMO}}{\lambda^{Mn}_{LVV}}} \,, \quad  I^{Mn}_{LMM} = \sum_{j=0}^{N} f_{Mn3+}*K*I_{Mn} e^{ \frac{-j*a_{LMO}}{\lambda^{Mn}_{LMM}}}, \quad\text{where} \quad f_{Mn3+} = \frac{1}{(N-q)(N-q-1)}
       \label{eq:6}
\end{equation} 

Here N = 7 and q =3, So $f_{Mn3+} = \frac{1}{(N-q)(N-q-1)} = \frac{1}{(7-3)(7-3-1)} = 0.0833$

\begin{equation}
      \frac{I^{Mn}_{LMM}}{I^{Mn}_{LVV}}  = \frac{ \sum_{j=0}^{19} f_{Mn3+}*K*I_{Mn} e^{ \frac{-j*a_{LMO}}{\lambda^{Mn}_{LMM}}}}{\sum_{j=0}^{19} I_{Mn} e^{ \frac{-j*a_{LMO}}{\lambda^{Mn}_{LVV}}}} \, \quad\quad   , \quad\text{where} f_{Mn3+} = 0.0833
       \label{eq:7}
\end{equation} 

Similarly if we have 20 layers of CMO with MnO$_2^{-1}$ termination and assuming all layers have Mn in the +4 oxidation state, we have

\begin{equation}
      \frac{I^{Mn}_{LMM}}{I^{Mn}_{LVV}}  = \frac{ \sum_{j=0}^{19} f_{Mn4+}*K*I_{Mn} e^{ \frac{-j*a_{CMO}}{\lambda^{Mn}_{LMM}}}}{\sum_{j=0}^{19} I_{Mn} e^{ \frac{-j*a_{CMO}}{\lambda^{Mn}_{LVV}}}} \,,  \quad\text{where} \quad f_{Mn4+} = \frac{1}{(7-4)(7-4-1)} =  0.1667
       \label{eq:8}
\end{equation} 

We are assuming the proportionality constant K is the same for all Mn oxidation states. 

In the case of charge transfer of LMO on STO and if we are looking at the ratio of the intensities of the Mn LMM and LMV Auger electrons, the intensity ratio at the Nth layer $\frac{I^{Mn}_{LMM}}{I^{Mn}_{LVV}}(N)$ with $N \geq 3$ would be given by,

\begin{equation}
      \frac{I^{Mn}_{LMM}}{I^{Mn}_{LVV}}(N)  = \frac{  f_{Mn2+}*K*I_{Mn}e^{ \frac{-(N-1)*a_{LMO}}{\lambda^{Mn}_{LMM}}} + \sum_{j=0}^{N-2} f_{Mn3+}*K*I_{Mn} e^{ \frac{-j*a_{LMO}}{\lambda^{Mn}_{LMM}}}}{I_{Mn}e^{ \frac{-(N-1)*a_{LMO}}{\lambda^{Mn}_{LVV}}} +\sum_{j=0}^{N-2} I_{Mn} e^{ \frac{-j*a_{LMO}}{\lambda^{Mn}_{LVV}}}} \,
       \label{eq:9}
\end{equation} 

Here, $f_{Mn2+} = \frac{1}{(7-2)(7-2-1)} =  0.05 \quad \text{and} \quad f_{Mn3+} = 0.0833 $
\par
In Equation \ref{eq:9}, we are assuming 1 layer of LMO close to the interface with Mn in the +2 oxidation state and the rest of the layers have Mn in the +3 oxidation states.

When we normalize with respect to the intensity ratio of the first layer, the proportionality constants K and $I_{Mn}$ disappear as shown below

\begin{equation}
     \frac{\frac{I^{Mn}_{LMM}}{I^{Mn}_{LVV}}(N)} {\frac{I^{Mn}_{LMM}}{I^{Mn}_{LVV}}(1)}  = \frac{\frac{  f_{Mn2+}*K*I_{Mn}e^{ \frac{-(N-1)*a_{LMO}}{\lambda^{Mn}_{LMM}}} + \sum_{j=0}^{N-2} f_{Mn3+}*K*I_{Mn} e^{ \frac{-j*a_{LMO}}{\lambda^{Mn}_{LMM}}}}{I_{Mn}e^{ \frac{-(N-1)*a_{LMO}}{\lambda^{Mn}_{LVV}}} +\sum_{j=0}^{N-2} I_{Mn} e^{ \frac{-j*a_{LMO}}{\lambda^{Mn}_{LVV}}}}}{\frac{ f_{Mn3+}*K*I_{Mn}}{I_{Mn}}} \,
       \label{eq:9}
\end{equation} 

\begin{figure}[h]
 \centering\includegraphics[width=\linewidth]{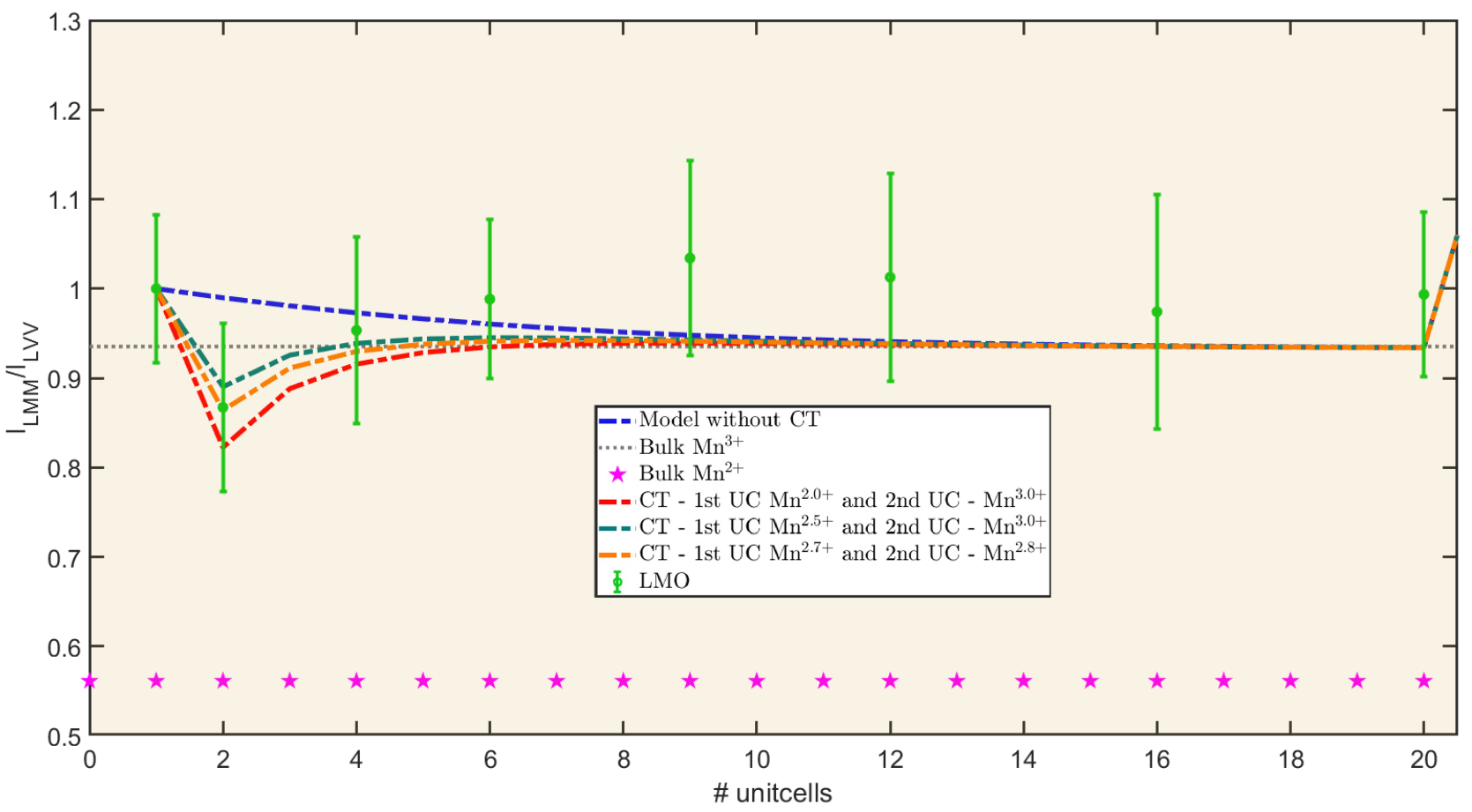}
 \captionof{figure}{LMM to LVV intensity ratios in 20 UC LMO on STO. Different charge distributions for the first two unitcells are modelled along with the experimental values (labelled as LMO). The Bulk +3 and +2 oxidation state levels of Mn are marked for reference.}
 \label{fig:LMOchargedist}
\end{figure}

\subsubsection{Vanadates}

\begin{figure}[h]
 \centering\includegraphics[width=\linewidth]{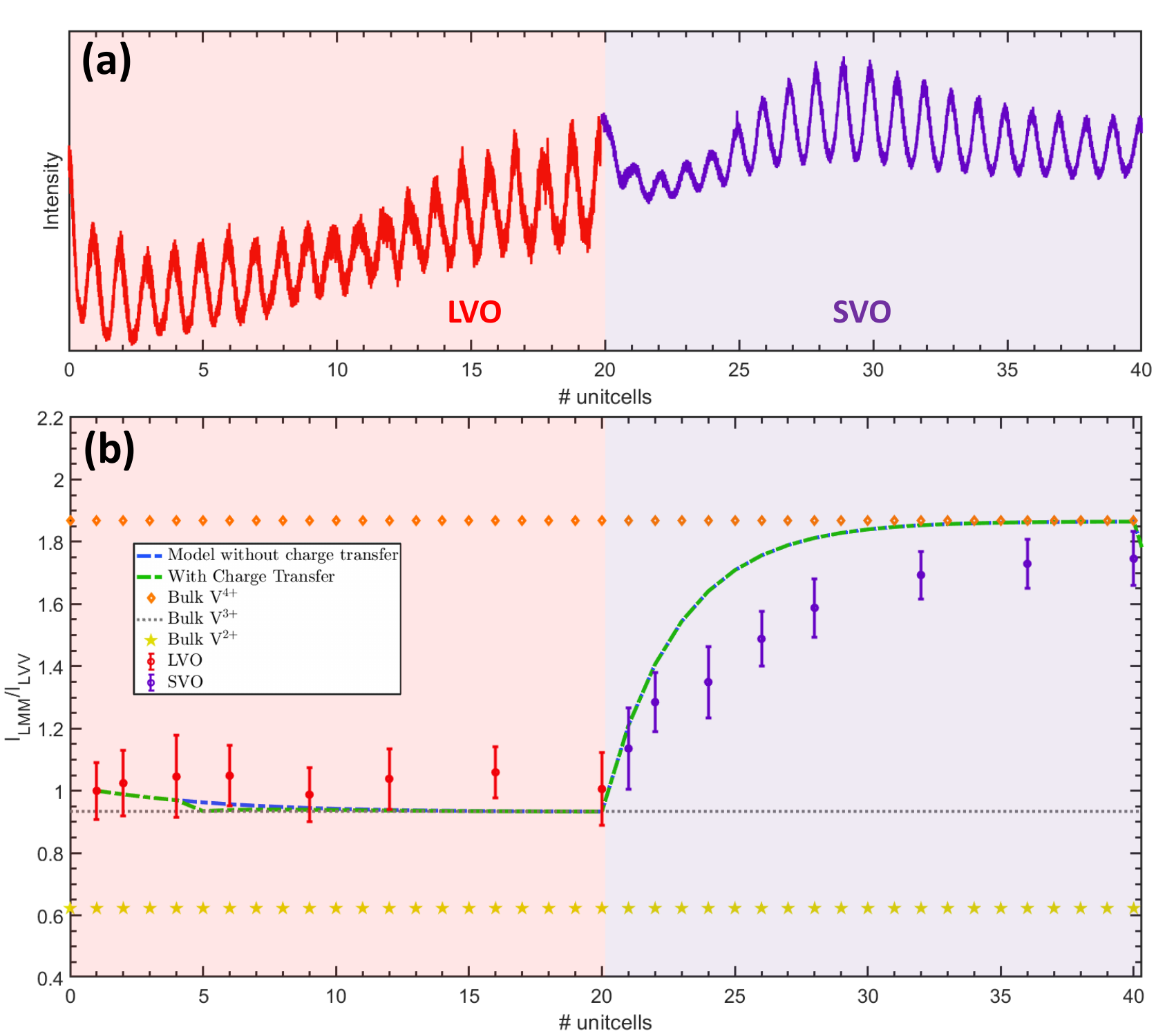}
 \captionof{figure}{(a)RHEED oscillations of LVO and SVO layers. (b)LMM to LMV intensity ratios in 20 UC SVO/ 20 UC LVO on STO. The Bulk +4,+3 and +2 oxidation state levels of V are marked for reference. Parameter free escape depth models with and without charge transfer in LVO/STO interface are plotted. The red and purple shaded regions represent LVO and SVO respectively.}
 \label{fig:LVOSVO_CT}
\end{figure}



\clearpage

\bibliographystyle{unsrt}
\renewcommand{\bibfont}{\small}
\bibliography{thesis}

\end{document}